\documentclass[12pt,aps,showpacs]{revtex4}%
\usepackage{amssymb}
\usepackage{amsmath}
\usepackage{amsfonts}
\usepackage{amsthm}
\usepackage{graphicx}%
\providecommand{\U}[1]{\protect\rule{.1in}{.1in}}
\newtheorem*{theorema}{Theorem}
\begin{document}
\title{Bertotti-Robinson and Melvin Spacetimes}
\author{David Garfinkle$^{1,2}$ and E.N. Glass$^{2}$}
\affiliation{$^{1}$Physics Department, Oakland University, Rochester, MI, $^{2}$Physics
Department, University of Michigan, Ann Arbor, MI}

\begin{abstract}
Similarities between the Melvin and Bertotti-Robinson spacetimes are discussed
and a uniqueness conjecture is formulated.

\end{abstract}

\maketitle

\section{INTRODUCTION}

The Bertotti-Robinson (BR) and Melvin (ML) spacetimes have interesting
similarities. Both are static, axisymmetric solutions of the Einstein-Maxwell
equations. Each solution has one free parameter which characterizes the
strength of the electromagnetic field. In this paper we will concentrate on
one particular property: that both spacetimes are static Einstein-Maxwell
solutions that are geodesically complete. To see why this is striking,
consider the analogous case of static, geodesically complete solutions of the
vacuum Einstein equation. A theorem of Lichnerowicz \cite{Bru62} states: The
only geodesically complete stationary vacuum spacetime (M,g) which is
asymptotically flat is Minkowski spacetime (R$^{4}$,$\eta$). Anderson
\cite{And04,And00} has considerably strengthened this result by allowing one
to deduce asymptotic flatness, not require it. He has generalized
Lichnerowicz's theorem as follows:

\begin{theorema}
[Anderson]The only geodesically complete stationary vacuum spacetime (M,g) is
Minkowski spacetime (${R}^{4},\eta$).
\end{theorema}

To prove this generalization, Anderson makes use of the geometric properties
of the Ernst equations, satisfied by a stationary vacuum spacetime. Since
static electrovac spacetimes also satisfy Ernst equations, one might expect
that the results of Anderson could be generalized to the electrovac case.
However, Anderson also notes that the analog of his theorem for
Einstein-Maxwell solutions is false because the Melvin solution is a
counterexample. Furthermore, the method of proof \cite{And04,And00} cannot be
generalized to the electrovac case because the electrovac Ernst equations have
different geometric properties than the vacuum Ernst equations.

There remains the question of whether static, geodesically complete electrovac
solutions are numerous or rare. It is our opinion that such solutions are
rare. We make the following conjecture.

\textbf{CONJECTURE:} \textit{The only geodesically complete static
Einstein-Maxwell spacetimes are Melvin and Bertotti-Robinson}

We have not been able to prove this conjecture. However, the remainder of this
paper will be a plausibility argument for the conjecture. For simplicity, we
will confine ourselves to electrovac spacetimes that are axisymmetric as well
as static. In section II we will consider the vacuum case (i.e. the well known
Weyl solutions) and examine which properties of this class of solutions
prevent the non-flat ones from being geodesically complete. In section III we
consider the Einstein-Maxwell equations and examine the properties of the ML
and BR spacetimes that allow them to be geodesically complete. Section IV has
further discussion, and details of the ML and BR spacetimes are contained in Appendices.

\section{Vacuum Solutions}

A vacuum or electrovac static, axisymmetric metric can be written in the
Weyl-Levi-Civita (WLC) form:
\begin{equation}
d{s^{2}}=-e^{2U}dt^{2}+e^{-2U}[e^{2K}(d\rho^{2}+dz^{2})+\rho^{2}d\varphi^{2}]
\label{Weylmetric}%
\end{equation}
where $U$ and $K$ are functions of $\rho$ and $z$. In the vacuum case, $U$
satisfies the flat space Laplace equation \cite{SKM+03}
\begin{equation}
\frac{\partial^{2}U}{\partial\rho^{2}}+\frac{1}{\rho}\frac{\partial
U}{\partial\rho}+\frac{\partial^{2}U}{\partial z^{2}}=0 \label{laplace}%
\end{equation}
while $K$ is determined by $U$ through the following equations:
\begin{align}
{\frac{\partial K}{\partial\rho}}  &  =\rho\left[  {{\left(  {\frac{\partial
U}{\partial\rho}}\right)  }^{2}}-{{\left(  {\frac{\partial U}{\partial z}%
}\right)  }^{2}}\right]  ,\label{drhoK}\\
{\frac{\partial K}{\partial z}}  &  =2\rho{\frac{\partial U}{\partial\rho}%
}{\frac{\partial U}{\partial z}.} \label{dzK}%
\end{align}
The solutions of Eq.(\ref{laplace}) can be written in closed form. They are
linear combinations of singular solutions of the form
\begin{equation}
U={r^{-(\ell+1)}}P_{\ell}(\cos\theta) \label{singU}%
\end{equation}
and non-singular solutions of the form
\begin{equation}
U={r^{\ell}}P_{\ell}(\cos\theta). \label{nonsingU}%
\end{equation}
Here $r$ and $\theta$ are given by $r=\sqrt{\rho^{2}+z^{2}}$ and $\theta
={\tan^{-1}}(\rho/z)$. $P_{\ell}$ denotes the Legendre polynomial of order
$\ell$. It is not surprising that the singular $U$ of Eq.(\ref{singU}) gives
rise to a singular spacetime. But why does the non-singular $U$ of
Eq.(\ref{nonsingU}) also give rise to a singular spacetime? At first one might
suspect that perhaps the metric function $K$ is singular, but it is easy to
see that that is not the case. The $U$ solutions of Eq.(\ref{nonsingU}) are
polynomials in $\rho$ and $z$ from which it follows, using Eqs.(\ref{drhoK})
and (\ref{dzK}), that $K$ is also a polynomial in $\rho$ and $z$. Instead, the
answer is found by examining curvature invariants, in particular the
Kretschmann scalar. For a static vacuum spacetime we have
\begin{equation}
{C^{abcd}}{C_{abcd}}=8{E^{ab}}{E_{ab}}%
\end{equation}
Here $E_{ab}$ is the electric part of the Weyl tensor defined by
\begin{equation}
{E_{ac}}={C_{abcd}}{n^{b}}{n^{d}}%
\end{equation}
where $n^{a}$ is the unit vector in the direction of the static timelike
Killing field. From Eq.(\ref{Weylmetric}) and the fact that $E_{ab}$ is
trace-free, it then follows that
\begin{equation}
{C^{abcd}}{C_{abcd}}=4{e^{4(U-K)}}\left[  {{({E_{\rho\rho}}-{E_{zz}})}^{2}%
}+3{{({E_{\rho\rho}}+{E_{zz}})}^{2}}+4{{({E_{\rho z}})}^{2}}\right]  .
\label{Kscalar}%
\end{equation}
The electric part of the Weyl tensor can be expressed in terms of $U$ by
\begin{equation}
{E_{ab}}={D_{a}}{D_{b}}U+{D_{a}}U{D_{b}}U
\end{equation}
where $D_{a}$ is the derivative operator associated with the spatial part of
the metric. From Eq.(\ref{Kscalar}) it is clear that an unbounded Kretschmann
scalar can be caused by the quantity $U-K$ being unbounded above. Thus, in the
WLC metrics that come from the nonsingular $U$ of equation (\ref{nonsingU})
the curvature blows up \textquotedblleft at infinity\textquotedblright%
\ (\textit{i.e.} at large $\rho$ and $z$) because $e^{4(U-K)}$ blows up at
infinity. Strictly speaking, a blowup of curvature \textquotedblleft at
infinity\textquotedblright\ does not make a spacetime singular unless
geodesics can get \textquotedblleft to infinity\textquotedblright\ in a finite
affine parameter. However, since spatial distance in the $\rho$ or $z$
direction is determined by the quantity $e^{K-U}$ it is not surprising that
when this quantity goes to zero at infinity some geodesic can get there in
finite affine parameter.

\section{Einstein-Maxwell Solutions}

For magnetostatic axisymmetric solutions of the Einstein-Maxwell equations,
there is a metric of the WLC form of Eq.(\ref{Weylmetric}) together with a Maxwell
field. The field $F_{ab}$ has zero electric component, $F_{ab}n^{a}=0$, with
the magnetic component determined from a scalar potential $\psi$:
\begin{equation}
{F_{tb}^\ast}={\partial_{b}}\psi
\end{equation}
where $F_{ab}^{\ast}$ is the dual of $F_{ab}$. The metric function $U$ and the
scalar potential $\psi$ satisfy the following equations \cite{SKM+03}:
\begin{align}
\frac{\partial^{2}U}{\partial\rho^{2}}+\frac{1}{\rho}\frac{\partial
U}{\partial\rho}+\frac{\partial^{2}U}{\partial z^{2}}  &  ={e^{-2U}}\left[
{{\left(  {\frac{\partial\psi}{\partial\rho}}\right)  }^{2}}+{{\left(
{\frac{\partial\psi}{\partial z}}\right)  }^{2}}\right]  ,\\
\frac{\partial^{2}\psi}{\partial\rho^{2}}+\frac{1}{\rho}\frac{\partial\psi
}{\partial\rho}+\frac{\partial^{2}\psi}{\partial z^{2}}  &  =2{\frac{\partial
U}{\partial\rho}}{\frac{\partial\psi}{\partial\rho}}+2{\frac{\partial
U}{\partial z}}{\frac{\partial\psi}{\partial z}.}%
\end{align}
The metric function $K$ is determined by $U$ and $\psi$ from the following
equations:
\begin{align}
{\frac{\partial K}{\partial\rho}}  &  =\rho\left[  {{\left(  {\frac{\partial
U}{\partial\rho}}\right)  }^{2}}-{{\left(  {\frac{\partial U}{\partial z}%
}\right)  }^{2}}\right]  +\rho{e^{-2U}}\left[  {{\left(  {\frac{\partial\psi
}{\partial z}}\right)  }^{2}}-{{\left(  {\frac{\partial\psi}{\partial\rho}%
}\right)  }^{2}}\right]  ,\label{drhoKem}\\
{\frac{\partial K}{\partial z}}  &  =2\rho{\frac{\partial U}{\partial\rho}%
}{\frac{\partial U}{\partial z}}-2\rho{e^{-2U}}{\frac{\partial\psi}%
{\partial\rho}}{\frac{\partial\psi}{\partial z}.} \label{dzKem}%
\end{align}

The Melvin solution has functions
\begin{align}
U  &  =\ln\left(  1+{{\frac{1}{4}}}{B_{0}^{2}}{\rho^{2}}\right) \\
\psi &  ={B_{0}}z\\
K  &  =2\ln\left(  1+{{\frac{1}{4}}}{B_{0}^{2}}{\rho^{2}}\right)
\end{align}
where $B_{0}$ is a constant representing the strength of the magnetic field.

At large $\rho$ we have $U\approx2\ln\rho$. However $U=2\ln\rho$ is actually a
vacuum solution, one of the well known Levi-Civita solutions. For this
solution $K=4\ln\rho$ and thus ${e^{U-K}}\rightarrow0$ as $\rho\rightarrow
\infty$. It then follows from Eq.(\ref{Kscalar}) that the Kretschmann scalar
also goes to zero as $\rho\rightarrow\infty$. However, unlike the ML solution,
the vacuum Levi-Civita solution is singular on the axis. Thus, the ML solution
seems to be a delicate compromise: at infinity it approaches a non-flat vacuum
solution, so $U$ must blow up at infinity. However, unlike the polynomial
solutions of Eq.(\ref{singU}) in the ML solution, $U$ blows up sufficiently
slowly at infinity so that the curvature does not blow up there. The presence
of the Maxwell field does not modify the behavior at infinity, but only serves
to make this behavior compatible with non-singular behavior on the axis. It is
the apparent delicacy of this compromise that leads to our conjecture.

We now turn to the Bertotti-Robinson solution to see whether it exhibits
similar behavior. For this solution we have
\begin{align}
U  &  =\ln\lambda+{{\frac{1}{2}}}\ln\left(  {\rho^{2}}+{z^{2}}\right) \\
\psi &  =\lambda{{\left(  {\rho^{2}}+{z^{2}}\right)  }^{1/2}}\\
K  &  =1
\end{align}
where $\lambda$ is a constant setting the magnetic field strength. At fixed
$z$ and large $\rho$ we have $U\rightarrow\ln\rho$. This is similar behavior
to what is seen in the ML case, though for BR the Maxwell field remains
constant and the solution does not approach any vacuum solution. Nonetheless,
the appearance of $\ln\rho$ behavior is striking: once again a singularity at
infinity is avoided by $U$ increasing at a rate smaller than polynomial. The
fact that both the ML and BR solutions do this, and that some property like
this seems to be needed to avoid singular spacetime behavior at infinity,
leads us to believe that precisely this logarithmic behavior is necessary.

\section{Discussion}

We have presented a uniqueness conjecture for the ML and BR spacetimes, and a
plausibility argument for that conjecture. However, a plausibility argument is
not a proof. It may be that the methods of Anderson \cite{And04,And00} could
be modified in some way to provide such a proof. On the other hand, it may be
that the conjecture could be falsified by finding a counterexample. The
strongest part of our plausibility argument is for the logarithmic behavior of
solutions at infinity where $U_{\text{ML}}\rightarrow\ln\rho^{2}$ and
$U_{\text{BR}}\rightarrow\ln\rho$; but possibly other solutions share this
behavior. One could search for such solutions numerically by using the methods
of Headrick et al \cite{HKW10}.

\section*{Acknowledgement}

DG was supported by NSF Grant PHY-0855532 to Oakland University.

\appendix

\section{Melvin Spacetime}

The ML metric \cite{Mel64} describes a static magnetic field which is a bundle
of magnetic flux lines in magnetostatic-gravitational equilibrium. Its line
element is given by
\begin{equation}
d{s^{2}}=a_{B}^{2}[- dt^{2}+d\rho^{2} + dz^{2} ] + a_{B}^{-2}\rho^{2}%
d\varphi^{2} \label{MLmetric}%
\end{equation}
with $a_{B}=1+B_{0}^{2}\rho^{2}/4$. The ML manifold has four Killing vectors
$\partial_{t},\partial_{z},\partial_{\varphi}$ and $z\partial_{t}%
+t\partial_{z}$. These Killing vectors are respectively time translation,
spatial translation along the axis, rotation around the axis, and boost along
the axis.

The Maxwell field is given by
\begin{equation}
F=B_{0}\rho a_{B}^{-2}\ d\rho\wedge d\varphi
\end{equation}
where $B_{0}$ is the value of the magnetic field on the $\rho=0$ axis. As with
any Einstein-Maxwell solution there are additional solutions given by duality
rotation: that is, the metric is unchanged but the Maxwell tensor $F_{ab}$
maps to ${F_{ab}}\cos\beta+{F_{ab}^{\ast}}\sin\beta$ where $\beta$ is any
constant and $F_{ab}^{\ast}$ is the dual of $F_{ab}$.

The ML metric as given in Eq.(\ref{MLmetric}) is already in the WLC form of
Eq.(\ref{Weylmetric}). Thus, we can immediately read off that
\begin{align}
U  &  =\ln a_{B}\\
K  &  =\ln a_{B}^{2}%
\end{align}
The Maxwell invariants are
\begin{align}
I_{1}  &  =\frac{1}{2}F_{ab}F^{ab}=B_{0}^{2}a_{B}^{-4}\\
I_{2}  &  =\frac{1}{2}F_{ab}^{\ast}F^{ab}=0
\end{align}
The ML metric is Petrov type D with the only non-zero Weyl tensor component
\begin{equation}
\Psi_{2}=-B_{0}^{2}a_{B}^{-4}(1-a_{B}/2).
\end{equation}
The Kretschmann scalar is
\begin{equation}
R_{abcd}R^{abcd}=4B_{0}^{4}a_{B}^{-8}[2+3(1-B_{0}^{2}\rho^{2}/4)^{2}]
\end{equation}
while a similar scalar involving the Ricci tensor is
\begin{equation}
{R_{ab}}{R^{ab}}=4B_{0}^{4}a_{B}^{-8}%
\end{equation}
Note that both of these scalars vanish as $\rho\rightarrow\infty$.

Melvin and Wallingford \cite{MW66} have computed all the geodesics paths in the ML spacetime,
and it follows from their computation that ML is geodesically complete.

\section{Bertotti-Robinson spacetime}

The BR spacetime \cite{LC,Ber59,Rob59,Dol68} has line element
\begin{equation}
d{s^{2}}=(\frac{1}{\lambda^{2}r^{2}})(-d\tau^{2}+dr^{2}+r^{2}d\vartheta
^{2}+r^{2}\sin^{2}\vartheta d{\varphi}^{2})\label{cf-br-met}%
\end{equation}
where $\lambda$ is a constant. This spacetime is thus the direct product of
the two-sphere and two dimensional anti de-Sitter spacetime. It therefore
inherits the symmetries of both these spaces and has a six parameter isometry
group. Since the line element is conformally flat, it follows that the Weyl
tensor is zero. Furthermore, the large isometry group insures that all
curvature scalars are constants.

Metric (\ref{cf-br-met}) can be put in WLC form by the coordinate
transformation
\begin{equation}
\tau=\lambda^{2}t,\text{ \ }\frac{\sin\vartheta}{r}=\rho,\text{ \ }\frac
{\cos\vartheta}{r}=z
\end{equation}
which yields the line element
\begin{equation}
d{s^{2}}=-{\lambda^{2}}({\rho^{2}}+{z^{2}})d{t^{2}}+{\frac{1}{{\lambda^{2}%
}({\rho^{2}}+{z^{2}})}}(d{\rho^{2}}+d{z^{2}}+{\rho^{2}}d{\varphi^{2}})
\label{br-wlc-met}%
\end{equation}
Comparing (\ref{br-wlc-met}) with metric (\ref{Weylmetric}) provides%
\begin{align}
U  &  =\ln\lambda+{{\frac{1}{2}}}\ln({\rho^{2}}+{z^{2}})\\
K  &  =1
\end{align}

The Maxwell field is given by
\begin{equation}
F={\frac{\rho}{\lambda{{({\rho^{2}}+{z^{2}})}^{3/2}}}}(zd\rho-\rho dz)\wedge
d\varphi
\end{equation}
The Maxwell invariants are
\begin{align}
I_{1} &  =\frac{1}{2}F_{ab}F^{ab}={\lambda^{2}}\\
I_{2} &  =\frac{1}{2}F_{ab}^{\ast}F^{ab}=0
\end{align}
The Kretschmann scalar is
\begin{equation}
R_{abcd}R^{abcd}=8{\lambda^{4}}%
\end{equation}
while a similar scalar involving the Ricci tensor is
\begin{equation}
{R_{ab}}{R^{ab}}=4{\lambda^{4}}%
\end{equation}

The geometric structure of BR is $S^{2}\otimes AdS_{2}$.  Therefore
(as pointed out {\it e.g.} by Cl\'{e}ment and Gal'tsov \cite{CG01}) since each of these two dimensional
spaces is geodesically complete it follows that BR is geodesically complete.

\end{document}